# 360 Virtual Reality Travel Media for Elderly

Donlaporn Srifar[1]

[1]Faculty of Mass Communication Technology, Rajamangala University of Technology Phra Nakhon, Bangkok, Thailand
donlaporn.sr@rmutp.ac.th

## Abstract

*The objectives of this qualitative research were to study the model of 360° virtual reality travel media, to compare appropriateness of moving 360° virtual reality travel media for elderly with both still and moving cameras, and to study satisfaction of elderly in 360° virtual reality travel media. The informants are 10 elders with age above and equal to 60 years old who live in Bangkok regardless of genders. Data were collected through documents, detailed interview, and non-participant observation of elders to 360° virtual reality travel media with data triangulation.*

*1. From the literature review on appropriate 360° virtual reality travel media design for elders, the results found that 1. The creation of 360° virtual reality must primarily consider the target consumers on their physics, 2. 360° virtual reality must have fluidity on changing view of the camera by calibrating with the target consumers, 3. The image displayed must not move too fast to prevent dizziness and improve comfort of the target consumers. It is also highly recommended to implement a function to customize movement rate for the customer, 4. The resolution of the display should be high and if the target consumers have visibility problems, they would use eyeglasses too to achieve best immersion, 5. The quality of sound effects and soundtracks must be high and clear enough. The sound should be stereo for the target consumers to recognize the direction of the sound.*

*2. From the in-depth interview of the target consumers, the results found that 1. They are worried and not used to the equipment, but they are interest in the equipment and did not feel fear, 2. They have no idea where to look, 3. They are dizzy, 4. They feel excited, 5. They are interest in what is more to see, 6 They feel like they did actually travel there, 7. They can hear sound clearly, 8. They do not like when the camera is moving and find still camera more comfortable.*

*3. From the non-participant observation, the researchers continuously observed the target consumers and found that they are always excited, laughed, and smiled when watching the media. They always asked where this is and why they cannot see anything when turning around. Some participants*

*cannot hear the sound clearly and some even have nausea like when they have got car sickness so they have to remove the equipment and rest before trying again. From the comparison, there are 9 out of 10 elders who have nausea from watching moving 360° virtual reality media but there is only 1 person who got nausea when watching still 360° virtual reality media.*

## KEYWORDS

*Virtual Reality, Elderly, Travel, Thailand, Koh Rattakosin, Temple*

## 1. INTRODUCTION

Currently, Thai society has a high tendency to become an aging society not so long from now according to the statistic of the growth in population of elderly people and the decrease of new children population. It has been predicted that in the near future around 10-20 years, we will live in a society full of elderly people. Other than the care for elders, one of things that require high attention is their mental health. Elders in the past period of Thailand would be in the care of their family and descendants so they would not feel lonely, but because of the changes in the structure of Thai society in the current date, the size of family has decreased from extended family, so the elderly people are left without anyone to take care of and feel lonely. They cannot travel alone because of their physical health or unable to be in crowded places. Therefore, they would live their late life with loneliness and feel that they have no worth and have no reason to live any longer. Humans, regardless if young or old, have their own value and are all worth living. Other than the elders who cannot travel alone, there are another group of people who also cannot travel due to their inability, which is the group of patients who cannot move themselves. These people are also in the risk of depression and feel like they have no worth.

Thailand is becoming an "Aging Society"[1] since 2005 according to the international criteria, which that country has the ratio of people age above 60 years old more than 10 percent or has the ratio of people age above 65 years old more than 7 percent of the total population. In year 2017, the number of elders in Thailand has already exceeded 17 percent of the population, which is more than 10 million people from the total population of 65 million people. These 10 million elderly people are in the state of dependency due to their physical health which have been decayed from aging and illness that would cause disability, such as a stork or accidents that would cause paralysis. From the intermittently survey, there is about 10 percent of 10 million elderly people, or one million elderly people, who are in the state of dependency. In this number, there are partial dependency people about 85 percent or about 850 thousand people who are "homestuck" and there are full dependency people about 15 percent or about 150 thousand people who are "bedstuck". The ratio of elderly who are homestuck or bedstuck is increasing from the increase of elderly people and the ratio of "ending age" that is also increasing as the average life expectancy of Thai people has increased respectively.

Thailand has described the definition of elderly as people who have their age more than 60 years (while most developed countries the age at 65 years old )according to the Act on Older Persons, B.E. 2546 (2003). The age range of elderly people has been divided into 3 major groups, which are the beginning age (60-69 years old), middle age (70-79 years old), and ending age (80 years old and above). The elderly people who are at their ending age will have more possibility to be homestuck or bedstuck than the beginning and the middle age.

The problem of homestuck and bedstuck of elderly people is a major problem for all countries including Thailand. Thailand has paid attention to the systematic solution the problems of elderly for a long time by having the first long-term national elderly plan with the period of 20 years since year 2003 in the government of Gen. Prem Tinsulanonda. Thailand currently using the second plan (2002-2021) and has a specific law for the elders, which is the Act on Older Persons, B.E. 2546 (2003). However, there are still many elderly people who are either homestuck or bedstuck.[1]

From the information above, it could be seen that the number of elders and patients who are unable to move is rather high. Therefore, if the people who are normally unable to travel outside could receive the same happiness like those who could, it might improve their mental health quality. In the order the achieve that, virtual reality technology is required by collecting the data from real locations for them, which is also called VR 360° (Virtual Reality 360°), which is a very popular technology by the current date and has a tendency to be developed and used more for the life of people. The VR technology works by collecting the pictures of surrounding environment with a special camera with 360-degree lense and display in special VR 360° equipment and the user will feel like they are actually participating in that displayed event.

The VR 360° or Virtual Technology 360° is a technology invented to virtualize the environment from both reality and imagination with computer technology and cooperation of over 200,000 developers and 700 startup companies across the world, and also the support of top companies in many industries which are directly related to the consumers including social network tops like Facebook, top technology company like Google, or even electronic appliance companies like Samsung and Sony. These companies paid close attention to VR technologies and created their own VR equipment such as Google Cardboard and Samsung Gear VR to support this technology. 360° Virtual Reality Technology was under the development since in the past but because of high cost of production, it was only limited to certain areas such as in medical area or in military area. When the time has passed, VR technology is more widespread and caused the reduction of production cost, which also make the companies in other countries use it as their marketing campaign to create new experience and thrilling for their customer such as Topshop, a fashion clothing brand, that give their customer new experience of watching fashion shows through their VR equipment from their home like they have actually watching the actual show.

Therefore, the researcher has interested in using VR 360° technology (Virtual Reality 360°) as a tool to create happiness for the elderly people and patients who are unable to move outside of their place[3], which leads this study of "360° Virtual Reality Travel Media for Elderly"

## 2. Objectives

2.1. To study the model of 360° virtual reality travel media

2.2. To compare appropriateness of moving 360° virtual reality travel media for elderly with both still and moving cameras

2.3. To study satisfaction of elderly in 360° virtual reality travel media.

## 3. Operational definition

The study "360° Virtual Reality Travel Media for Elderly", which is a qualitative research by collecting data from the observation and in-depth interview to collect data with non-structured interview with the objective of studying the satisfaction of elderly people to 360° Virtual Reality Travel Media.

## 4. Methodology

### 4.1 Target Group and Material

4.1.1. Target: The targets will be selected and sorted from the Foundation for Older Persons' Development 2017. 10 elderly people age more than 60 years old who live in Bangkok regardless of genders are selected with purposive sampling.

4.1.2. Materials for data collection:

− Field Note: A note for recording the content and details of the certain time.

− Interview: This study uses non-structured interview, which the interviewer will not ask the strict questions and will not ask orderly, and even ask something else outside the main questions according to the situation but will not exceed the study range.

- Non-Participant Observation: The researchers will observe the participants while they are watching 360° Virtual Reality Travel Media without participating.

4.1.3. Qualitative Data Collection: The researchers have collected data since August 2017 to March 2018 as followed.

- Study related documents and research: Study the concept of 360° virtual reality media, traveling concept, the concept of the behavior of Thai elderly people, recognition concept, and research about the mental health of elderly people.

- Detailed Interview: Use non-structured interview, which the interviewer will not ask the strict questions and will not ask orderly, and even ask something else outside the main questions according to the situation but will not exceed the study range.

- Non-Participant Observation: The researchers will observe the participants while they are watching 360° Virtual Reality Travel Media without participating

# Research Methodology

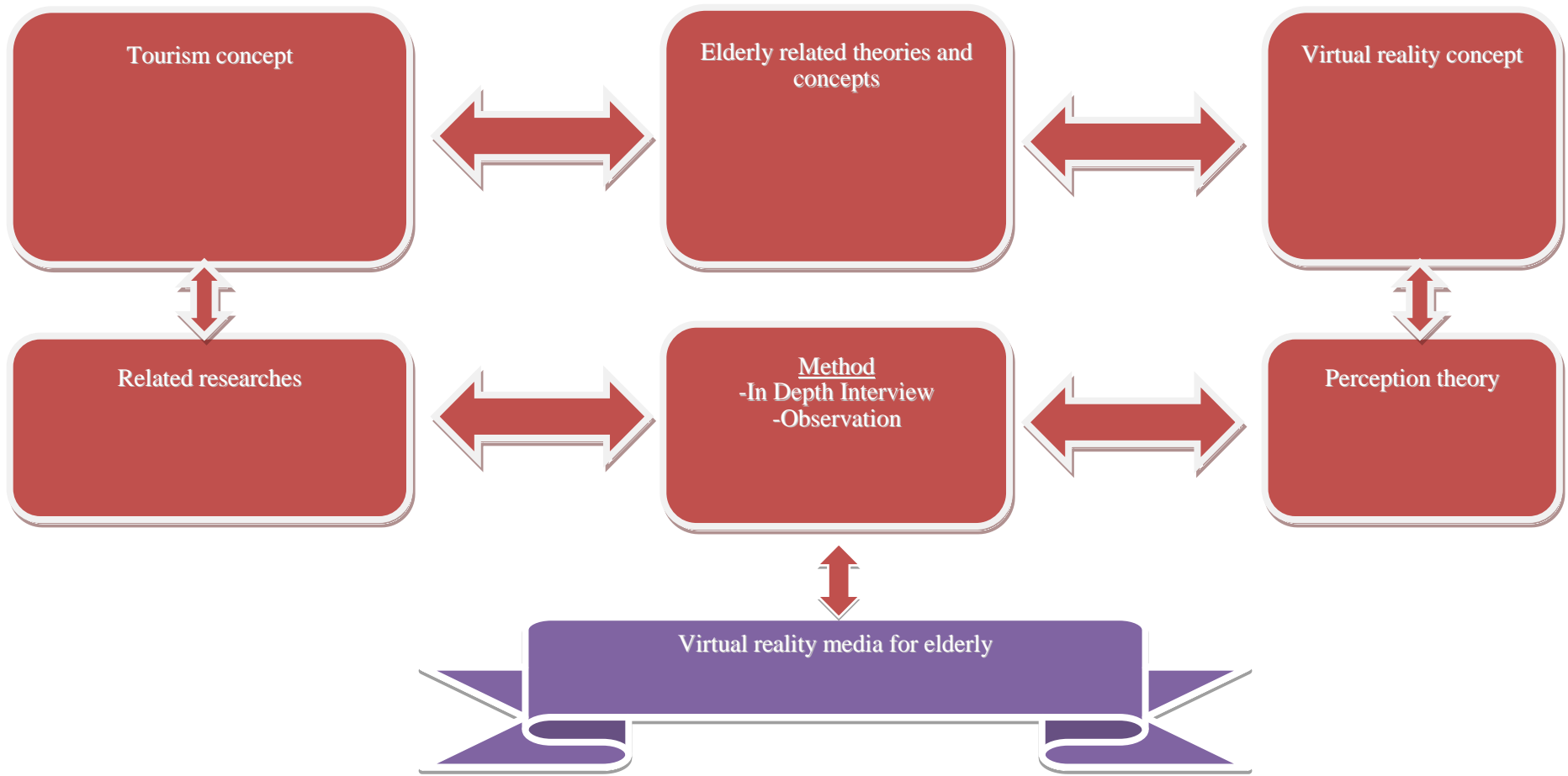

picture 1. Research Methodology.

# 5. Results

**Chapter 1: The appropriate form of 360° virtual reality travel media for the elderly**

From the qualitative analyzation of documents and interview, we have found that:

1. The creation of 360° virtual reality must primarily consider the target consumers on their physics.
2. 360° virtual reality must have fluidity on changing view of the camera by calibrating with the target consumers.
3. The image displayed must not move too fast to prevent dizziness and improve comfort of the target consumers. It is also highly recommended to implement a function to customize movement rate for the customer.
4. The resolution of the display should be high and if the target consumers have visibility problems, they would use eyeglasses too to achieve best experience.
5. The quality of sound effects and soundtracks must be high and clear enough. The sound should be stereo for the target consumers to recognize the direction of the sound.

**Chapter 2: Interview Observation**

From the interview of the elderly, we have found that most of them have eyesight problems and some have the problem of hearing, they have dizziness and could not watch moving 360° virtual reality travel media, have worries, and not used to the equipment, but they are interest in the equipment and did not feel fear and instead feel excited. They are interest in what is more to see, feeling like they did actually travel there. They can hear sound clearly but they do not like when the camera is moving and find still camera more comfortable.

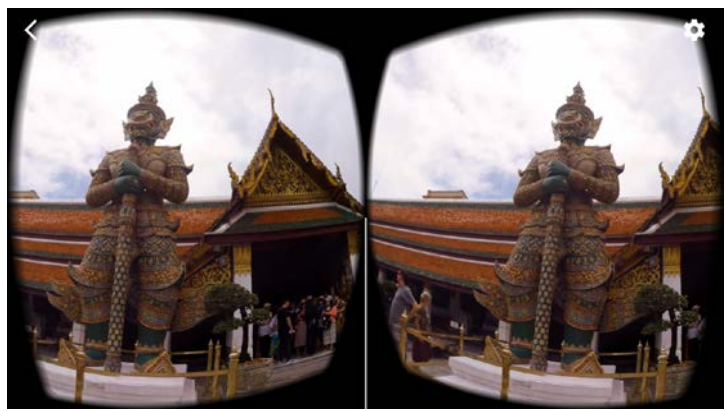

picture 2. VR 360° Giant of Grand Palace.

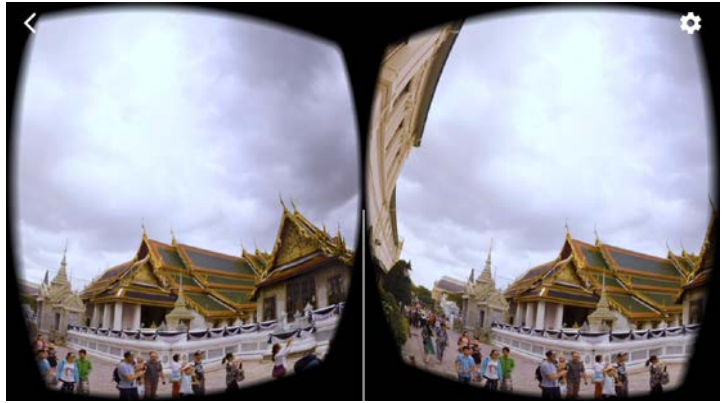

picture 3. VR 360° Grand Palace.

*"I feel like I'm having a car sick, I have no idea where it will move to too"*

*"I can't see well. This is a place to pay the homage but I can't see the prayers. Can I wear glasses?"*

*"Why is it rotating? Where should I start looking?"*

*"The texts are so small, it should be bigger"*

*"I feel dizzy. Where should I move my head to? It's feel like it's tilted"*

When the elderly people switched to watch 360° virtual reality travel media which is not moving, they felt relaxed and could move their head around according to the narration.

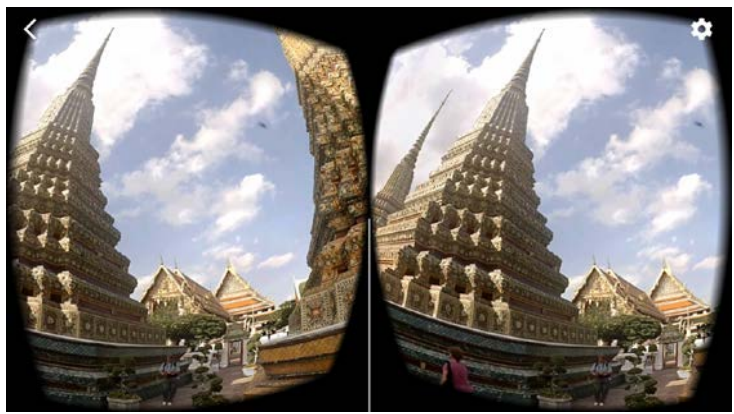

picture 4. VR 360° Wat Arun.

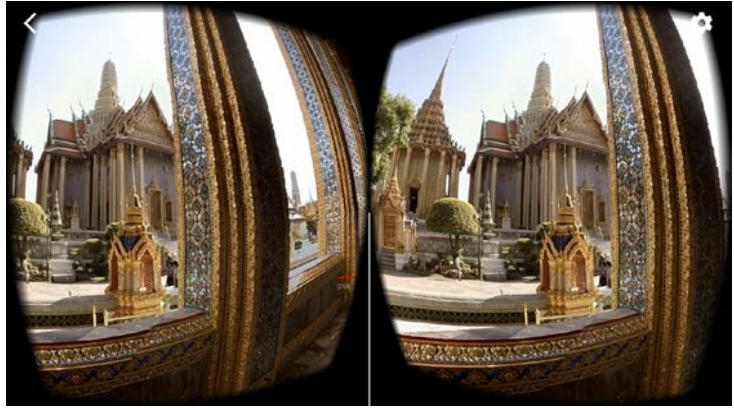

picture 5.  VR 360° Grand Palace.

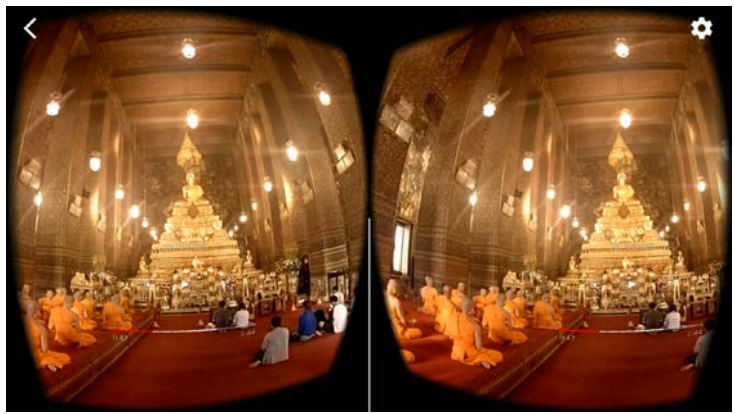

picture 6.  VR 360° In side of Chapel of Buddha.

"Is that the city hall?"

"Oops! It's already changed to Wat Chana Songkhram?"

"It's done here, where will you take me next?"

"Why don't you bring me to other places too? Like a sea or a mountain. I want to visit Phu Kradueng too"

"This is so much better. I can look around without any dizziness"

"No more dizziness. It's still and changing places. Good"

"Now I can follow the sound better, unlike before"

**Chapter 3: The satisfaction of the elderly people in 360° virtual reality travel media**

From non-participant observation, which the researchers have always observed the participants while they are watching the media, the participants are always excited, laughed, and smiled when watching the media. They always asked where is this and why they cannot see anything when turning around. Some participants cannot hear the sound clearly and some even have nausea like when they have got car sickness so they have to remove the equipment and rest before trying again. From the comparison, there are 9 out of 10 elders who have nausea from watching moving 360° virtual reality media but there is only 1 person who got nausea when watching still 360° virtual reality media.

From the analyzation of qualitative data from the interview, elderly who have watched 360° virtual reality travel media have enjoyed the media, gain new experience, feel like they have actually travelled there, happy, but the dizziness made some of them not wanting to watch the media again. However, almost all of them are interested and want to travel other places too.

## 6. Discussions

According to this research of the use of 360° virtual reality travel media, to make the target group able to enjoy it deeply, there is a need to consider the experience of their life in the making too such as their living environment, country, province, and their native language to make them feel familiar and could enjoy the experience without any boredom, which is in accordance to the research of Jose Luis Rubio-Tamayo, ID, Manuel Gertrudix Barrio 2 ID, and Francisco Garcia Garcia (2017) [4] which is studied on the topic of Immersive Environments and Virtual Reality: Systematic Review and Advances in Communication, Interaction, and Simulation.

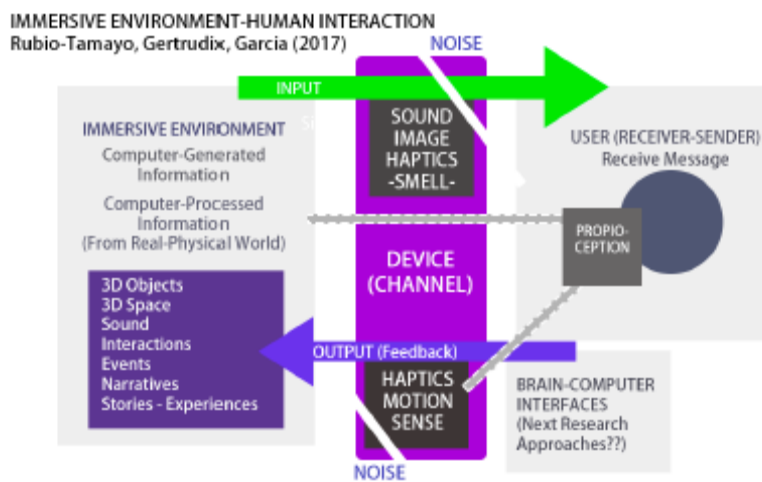

picture 2. Immersive Environment-Human Interaction
(Rubio-Tamayo, Gertrudix and Garcia (2017)).

It is also according to Maffei, L.; Iachini, T.; Masullo, M.; Aletta, F.; Sorrentino, F.; Senese, V.P.; and Ruotolo, F.; which studied on Individual reactions to a multisensory immersive virtual environment: the impact of a wind farm on individuals[5], which is also agreeable with Ruotolo, F.; Maffei, L.; Di Gabriele, M.; Iachini, T.; Masullo, M.; Ruggiero, G.; Senese, V.P.[6] that have found that the use of clear and appropriate sound with the elements in the creation of immersion will help the target group enjoy their vistual more,[6] which this research too is also agreeable with the research of Cho, S.; Ku, J.; Cho, Y.K.; Kim, I.Y.; Kang, Y.J.; Jang, D.P.; Kim, S.I.[7] that the use of virtual reality in rehabilitation therapy for the stroke patients could help them, but it would take more than 3 months. The research is also according to the research of Serrano, B.; Baños, R.M.; Botella, C. [8] that smell could enticing the relaxation of the target for more enjoyable and immersive experience.

## 7. Practical Suggestions

The things should be considered in the making of 360° virtual reality travel media for the elderly is to primarily consider the target consumers on their physics, the image displayed must not move too fast to prevent dizziness and improve comfort of the target consumers, make the target feel relaxed when watching this media; it is also highly recommended to implement a function to customize movement rate for the consumers in order to not make them feel dizzy. The resolution of the display should be high and if the target consumers have visibility problems, they would use eyeglasses too to achieve best immersion. The quality of sound effects and soundtracks must be high and clear enough. The sound should be stereo for the target consumers to recognize the direction of the sound and to not confuse them.

## 7. References

[1]     Chompunuch Phromphak, (2014), "Aging society in Thailand." *Academic Office of the Secretary of the Senate*, Year 3, 16 August 2014.

[2]     Dr.wichai Shokwiwat, (2017), *Caring for the elderly homestuck or bedstuck.*, 2017. Post Today Newspaper, 7 March 2017.

[3]     Rosie Wolf Williams, (2017), "*Rosie Wolf Williams*", Next Avenue Contributor, MAR 14, 2017.

[4]     Jose Luis Rubio-Tamayo Manuel Gertrudix Barrio and Francisco García García ,(2017), "*Immersive Environments and Virtual Reality: Systematic Review and Advances in Communication, Interaction and Simulation*", Multimodal Technologies and Interaction, September 2017.

[5]     Ruotolo, F., Senese, V.P., Ruggiero, G., Maffei, L., Masullo, M., & Iachini, T. (2012). "*Individual reactions to a multisensory immersive virtual environment: the impact of a wind farm on individuals. Cognitive Processing*". 13, 319-323.

[6]     Ruotolo, F.; Maffei, L.; Di Gabriele, M.; Iachini, T.; Masullo, M.; Ruggiero, G.; Senese, V.P., (2015), "*On the Validity of Immersive Virtual Reality as Tool for Multisensory Evaluation of Urban Spaces*", 6th International Building Physics Conference, IBPC 2015.

[7]     Cho, Y.K.; Kim, I.Y.; Kang, Y.J.; Jang, D.P.; Kim, S.I., (2014), "*Development of virtual reality proprioceptive rehabilitation system for stroke patients.*" Computer Methods and Programs in Biomedicine. Volume 113, Issue 1, January 2014.

[8]     Serrano, B.; Baños, R.M.; Botella, C., (2015), *"Virtual reality exposure-based therapy for the treatment of post-traumatic stress disorder: a review of its efficacy, the adequacy of the treatment protocol, and its acceptability."* Dove Press journal: Neuropsychiatric Disease and Treatment, 7 April 2016.